\documentclass[10pt, conference]{IEEEtran}
\IEEEoverridecommandlockouts
\usepackage{cite}
\usepackage{amsmath,amssymb,amsfonts}
\usepackage{graphicx}
\usepackage{textcomp}
\usepackage{xcolor}

\usepackage{algorithm}
\usepackage{algpseudocode}

\usepackage{subfigure}

\algblock{Input}{EndInput}
\algnotext{EndInput}
\algblock{Output}{EndOutput}
\algnotext{EndOutput}

\def\BibTeX{{\rm B\kern-.05em{\sc i\kern-.025em b}\kern-.08em
    T\kern-.1667em\lower.7ex\hbox{E}\kern-.125emX}}
\begin{document}

\title{RPDP: An Efficient Data Placement based on Residual Performance for P2P Storage Systems}


\author{
\IEEEauthorblockN{
Fitrio Pakana\IEEEauthorrefmark{1}
Nasrin Sohrabi\IEEEauthorrefmark{2}, 
Chenhao Xu\IEEEauthorrefmark{2},  
Zahir Tari\IEEEauthorrefmark{2}, 
and Hai Dong\IEEEauthorrefmark{2}}

\IEEEauthorblockA{
\IEEEauthorrefmark{1}\IEEEauthorrefmark{2}School of Computing Technology \\
Centre of Cyber Security Research and Innovation (CCSRI)\\ RMIT University Melbourne, Australia\\
Email: 
\IEEEauthorrefmark{1}{\it s3778275@student.rmit.edu.au}
\IEEEauthorrefmark{2}\{nasrin.sohrabi, chenhao.xu, zahir.tari, hai.dong\}@rmit.edu.au.
}
}


\maketitle

\begin{abstract}

Storage systems using Peer-to-Peer (P2P) architecture are an alternative to the traditional client-server systems. They offer better scalability and fault tolerance while at the same time eliminate the single point of failure. The nature of P2P storage systems (which consist of heterogeneous nodes) introduce however data placement challenges that create implementation trade-offs (e.g., between performance and scalability). Existing Kademlia-based DHT data placement method stores data at closest node, where the distance is measured by bit-wise XOR operation between data and a given node. This approach is highly scalable because it does not require global knowledge for placing data nor for the data retrieval. It does not however consider the heterogeneous performance of the nodes, which can result in imbalanced resource usage affecting the overall latency of the system. Other works implement criteria-based selection that addresses heterogeneity of nodes, however often cause subsequent data retrieval to require global knowledge of where the data stored. This paper introduces Residual Performance-based Data Placement (RPDP), a novel data placement method based on dynamic temporal residual performance of data nodes. RPDP places data to most appropriate selected nodes based on their throughput and latency with the aim to achieve lower overall latency by balancing data distribution with respect to the individual performance of nodes. RPDP relies on Kademlia-based DHT with modified data structure to allow data subsequently retrieved without the need of global knowledge. The experimental results indicate that RPDP reduces the overall latency of the baseline Kademlia-based P2P storage system (by 4.87\%) and it also  reduces the variance of latency among the nodes, with minimal impact to the data retrieval complexity.

\end{abstract}


\section{Introduction}\label{sec:introduction}
The paradigm of leveraging combined storage capacity of peers in network, known as peer-to-peer (P2P) storage, has been around for decades. Emerges as the alternative to traditional client-server storage, it aims to provide a system that moves away from the need of central component, hence, eliminating the single point of failure and capable of storing data with better scalability and fault tolerance with its inherent local nature \cite{Kermarrec2015}. Along with the explosive growth of data, recent years have witnessed the birth of modern P2P storage systems and protocols such as the Interplanetary File System (IPFS), Storj, and Swarm among others.

Data placement method is one of the core design choices in any of P2P storage system that defines distinctive characteristic of the system in terms of performance, scalability, and reliability. Some problems often contributed by data placement method includes: 

\begin{itemize}
\item \textit{Imbalance resource usage}. Kamdelia-based DHT data placement method stores data to the node with ID closest to the ID of data using the seminal XOR-based distance approach \cite{Maymounkov2002}, achieving deterministic routing convergence for data store and retrieval with $\mathcal{O}(\log{n})$ complexity without requiring of global knowledge. This method does not however consider the  performance of heterogeneous nodes, and this can result in imbalance of resource usage and data distribution in respect to the individual performance of each node i.e., data can be stored at busy nodes or nodes with lower performance while leaving other non-busy nodes or nodes with higher performance under utilized \cite{Lu2022}. Both IPFS \cite{Daniel2022, Trautwein2022} and Swarm \cite{Swarm2021} have used such as method for data placement and retrieval. The imbalance resource usage leads to higher latency, wasted storage space, and diminished reliability \cite{Lu2022}. 

\item \textit{Reduced scalability}. Other data placement methods such as the criteria-based \cite{Storj2018, Lu2022, Zhou2019}, maintain a global view of status of the nodes and perform storage node selection based on this view. However, such methods often require the system to maintain a global map contains storage node location of the data to be used during data retrieval. Such a design introduces the need of central component during data retrieval, and therefore reduces the scalability of the system as the central component needs to be scaled as the number of nodes grows in the system. The newest version of Storj (V3) \cite{Storj2018} is using this method for its data placement and retrieval.  
\end{itemize}

To address the problems described above, one appropriate approach is to design a data placement method that is able to balance the workload of the nodes across the system in proportion to their performance. In relation to data storage, this means nodes with higher performance will be assigned more workload to store data compared to the lower ones. The method must also be designed in such way that stored data can be retrieved without requiring central component. This paper proposes Residual Performance-based Data Placement (RPDP), a novel data placement method based on dynamic residual performance of the nodes within P2P storage systems. \textit{Residual performance} is defined as the remaining performance capacity of a node to handle incoming workload, measured using throughput and latency. The objective of RPDP method is to achieve balanced workload with respect to the individual performance of nodes which in turn can improve the overall latency of the system. Further, the RPDP method is designed to allow corresponding data retrieval method to rely on Kademlia-based DHT protocol to retrieve the data, eliminating the need of central component and therefore minimizing the impact to scalability of the system.

To the best of our knowledge, RPDP is the first attempt to perform a criteria-based for data placement and Kademlia-based DHT for data retrieval. The specific contributions of this paper are summarized as follows:

\begin{itemize}
  \item A data placement method for P2P storage system that considers the dynamic performance of the  nodes in a given system. Specifically, each node periodically calculates its current residual performance based on current throughput and latency and provides the metrics to corresponding consolidator node (termed as monitor node, in this paper). The latter in turn provides a global view, a "score board"-like that ranks the performance of all nodes within its responsibility area (termed as cluster, in this paper). The monitor node selects node with highest rank and provide the information to the requesting client which then proceeds to store data into the node (termed as actual node). Subsequently, a client looks up for closest node to the data and stores a location map into the node (termed as virtual node). To accommodate such a design, RPDP modifies the DHT data structure to allow the storage of two types of data: the data itself and the location map.  
  \item A data retrieval method for P2P storage system that is adapted specifically to the modified DHT data structure. Using Kademlia protocol, given the ID of data, it looks up for closest node and based on the data location map stored in this node, it performs another look up to find at the actual location of the data.  
\end{itemize}

We conducted experiments where we compared RPDP with the state-of-the-art Kademlia-based DHT data placement method using overall latency and latency variance metrics. Evaluation results indicate that RPDP achieves lower overall latency (by 4.87\%) and less brittle latency variance indicating that workload distribution of nodes in RPDP is more balanced compared to the Kademlia-based DHT. We also provided complexity analysis of the corresponding data retrieval method and concluded that the additional complexities is subtle and the overall complexity remains the same as the Kademlia-based DHT.

The rest of this paper is organized as follows. Section \ref{sec:related_work} describes the related existing works while Section \ref{sec:background} elaborates the existing ones that are adopted in the implementation of our work. We introduce the design of RPDP data placement with corresponding data retrieval method in Section \ref{sec:design}. Section \ref{sec:evaluation} presents the experimental setup and analysis of the results. Finally, Section \ref{sec:conclusion} concludes the paper.

\section{Related Work}\label{sec:related_work}

This section describes some prominent existing work that uses Kademlia-based DHT data placement methods as well as those using the criteria-based, including their corresponding data retrieval methods. From the body of the literature, we found that most that uses criteria-based method designed not specifically for P2P storage systems but rather for a more generic distributed storage systems. Nonetheless, a description of the methods is provided. 

\subsection{Kademlia-based DHT Data Placement}

Prominent works that use Kademlia-based DHT data placement and/or retrieval method as part of their proposal and implementation for P2P storage systems includes IPFS, Swarm, SAFE, and AuroraFS. 

Interplanetary File System (IPFS) \cite{Benet2014, Trautwein2022} was introduced by Protocol Labs in 2015 and has gained popularity as storage layer for blockchains \cite{Daniel2022}. IPFS uses Kademlia-based DHT only for placement and replication of \textit{provider record}, a structure that contains mapping between content identifier to node identifier as indication that such node is hosting/storing the data. Data initially stored by the initiator node itself as a content provider along with corresponding provider record. The node makes this known to the peers by publishing the provider record into the Kademlia-based DHT. The peers that choose to store the data subsequently retrieve the data from the publisher, store the data, then publish their own provider record into the network. Data retrieval in IPFS is using Bitswap protocol, which primarily using the Kademlia-based DHT but in opportunistic fashion i.e., before entering the DHT look up, the requesting peer asks all peers whether they already stored the data. Introduced initially as native base layer of Ethereum, Swarm \cite{Swarm2020, Swarm2021} is now a distributed storage platform of its own. The data placement method in Swarm is using Kademlia-based DHT i.e., data is stored at the neighbourhood of closest nodes by measured by bitwise XOR-distance. The corresponding data retrieval is also using Kademlia-based DHT i.e., by routing the request towards the relevant neighbourhood which calculated based on the distance. Further, Swarm also implemented data caching scheme whereby nodes who participate in routing retrieval may choose to store the data they have forwarded. Secure Access For Everyone (SAFE) \cite{Safe2014, Safe2018} introduces a security layer on top of the Kademlia-based DHT data placement method that it uses. Data store at the closest node within a network section must be approved by special node called the elders which are the oldest nodes in the section. Its data retrieval also using Kademlia-based DHT with added security layer for private data i.e., clients need to to perform an authentication. AuroraFS \cite{AuroraFS2022} focuses on storage and streaming media, claimed to be the first P2P file system that implement authorised data access capabilities, therefore meeting industrial-grade needs. Selection of nodes to store the data is using Kademlia-based DHT. AuroraFS implements 'Ant Colony Algorithm' in its data retrieval method in which every node passed en route to the location of data chunk is given a calculated rank based on size of its data and number of read requests are currently served by the node. For subsequent requests, the data will be read from node with the highest ranking.

The IPFS data placement method only uses Kademlia-based DHT to store provider record while the location of node storing data is random, and therefore likely to result in imbalance resource usage. Swarm, SAFE, and AuroraFS are using Kademlia-based DHT data placement in a more purity form therefore by using  larger identifier (ID) space and randomly assigning ID to nodes, data distribution in the network likely to be more balanced compared to IPFS. However, none of these data placement methods considers the dynamic performance of nodes in selecting location to store data.

Dual use of a certain protocol for both data placement and data retrieval (data indexing/routing) e.g., the use of Kademlia-based DHT, can simplify the P2P system architecture and implementation. However, it is necessary to have clear distinction between the two to identify full spectrum of design space \cite{Datta2009}. Nonetheless, the design of data placement strategy normally leads to a corresponding, specific data retrieval strategy. In other words, the data retrieval strategy must be designed according to how data are stored.

\subsection{Criteria-based Data Placement}

Storj \cite{Storj2018, Sammy2021} is a decentralized cloud storage (DCS) that provides compatibility with Amazon S3 application programming and uses Reed Solomon erasure coding (EC) \cite{Reed1960} for better availability of files in the network. To store data, client contacts a special node called Uplink node which in turn connects to another special node called Satellite node. Satellite node is the one that performs selection of storage nodes based on their reputation. In the newest version of Storj (v3), factors that determine reputation of storage node includes the latency (measured via ping response time to the Satellite node), uptime, history of reliability, and other desirable qualities. To minimize the tail latency, the erasure coding configured to over-encode the stripes to more pieces than needed, providing redundancy. The method further requires Satellite nodes to maintain a meta-data that contains mapping between data and the storage nodes needed for data retrieval. Reinforcement Learning Replica Placement (RLRP) \cite{Lu2022} is another data placement method that can be considered as criteria-based. It uses data placement agent which constructed through the Deep-Q-Network (DQN) model \cite{Mnih2015}, one of the reinforcement learning (RL) algorithms. Given the current states and attributes of data nodes, the agent uses an algorithm to determine placement of data replicas to data nodes (“the action”) that maximizes the output value of a function (“the reward”). Reward is a scalar value which is the negative of standard deviation of current state of data nodes. After data replicas placed, the state of the nodes (“the state”) is updated and fed back to the agent for subsequent placement. RLRP maintains Replica Placement Mapping (RPM) which responsible for defining the mapping of data to the storing nodes.

The data placement method of both Storj and RLRP considers the dynamic performance of nodes when making selection of storage location. However, the systems need to maintain a global view containing mapping of data to the storing nodes to be used in their corresponding data retrieval, which introduces the need of central component and therefore reduces the scalability of the system.

Although only used for subsequent replica placement, Pattern-directed Replication Scheme (PRS) \cite{Zhou2019} method provides an approach to store data by performing matching between characteristics of data with the characteristics of the nodes. Specifically, PRS analyzes the data access pattern and heterogeneous characteristics of nodes in the system (e.g., capacity, bandwidth, etc) then store data into matching nodes. Similar with Storj and RLRP, this method also considers the dynamic performance of nodes, however its approach diminishes the scalability of the system and requires complex algorithm with high time and space complexity \cite{Lu2022}.  

\section{Background}\label{sec:background}
This section provides description of two existing works that are applied in the Peer-to-Peer (P2P) storage systems considered in our work, namely the Kademlia overlay network and Connectivity-based Distributed Node Clustering (CDC) clustering method.

\subsection{Kademlia Overlay Network}\label{subsec:bg_kademlia}

An overlay network is logical network that is laid on top of another network, typically constructed to support application-specific needs. One of the well-known overlay networks for P2P system is Kademlia, introduced in 2002 \cite{Maymounkov2002}. In Kademlia overlay network, each node connected to the other nodes (peers/neighbours) by the notion of distance measured using exclusive or (XOR) of the ID between two nodes. Typically, the node ID is a unique large number assigned when the node joins the network by running an identifier generator function, e.g., 160-bit SHA-1. Every node maintains a routing table which is a list (also known as $k$-bucket, value of $k$ is implementation-specific) of the necessary information to connect and communicate with their respective neighbours. Neighbours are peers having XOR-based distance as steps of $2^i$ further away from a node within $n$-bit identifier space, $0 \leq i < n$. As such, it can happen that two nodes having similar ID are considered as neighbours while physically, they are at different geographical location. One of the most advantageous characteristics of Kademlia overlay network is that its algorithm offers deterministic routing convergence with $\mathcal{O}(\log{n})$ complexity.

\subsection{CDC Network Clustering}\label{subsec:bg_introduction}

\begin{figure*}[!h]
\centerline{\includegraphics[width=0.7\textwidth]{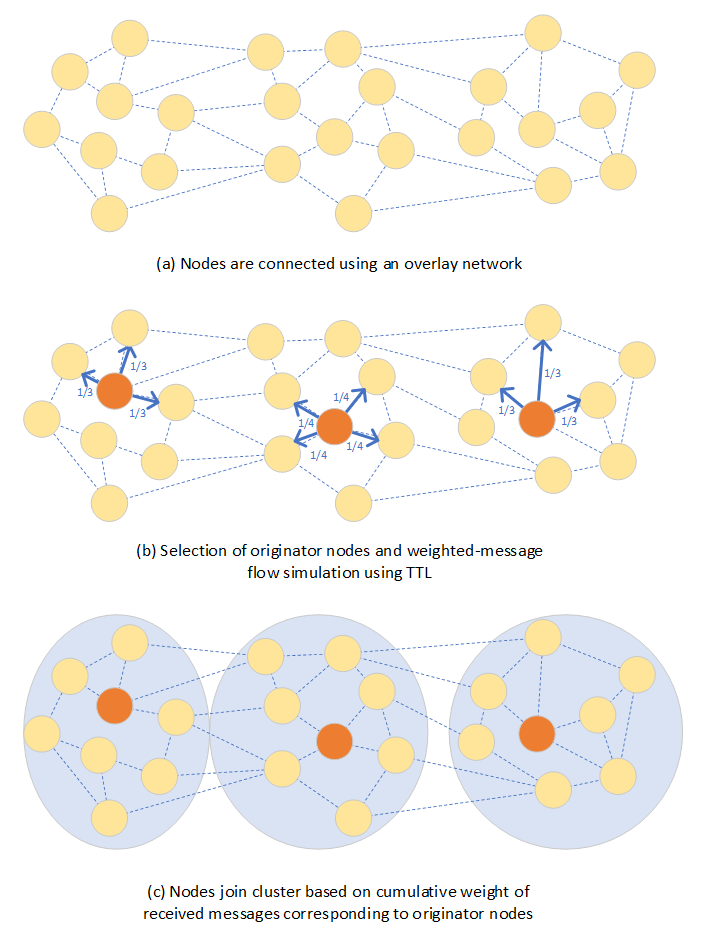}}
\caption{CDC Network Clustering}
\label{fig:clustering}
\end{figure*}

Network clustering involves arranging nodes in a network in such a way that nodes having similar characteristics are grouped together. It can be applied to overlay network, for example, to improve the routing efficiencies \cite{Singh2007}. One of the prominent clustering methods that is originally designed for P2P system is Connectivity-based Distributed Node Clustering (CDC) introduced in \cite{Ramaswamy2005}. This method is fully decentralized and does not require global knowledge of the network. Instead, it relies only on the local knowledge about neighbouring nodes to automatically cluster the entire network. Further, it has mechanism in handling of network churn (i.e., peers joining and leaving network) in such a way that re-clustering is not necessary if the number of nodes entering or exiting the network is not substantial. The CDC method clusters a P2P network represented as an unweighted and undirected graph $G = \left(V, E\right)$ where any vertex $v \in V$ represents a node in the network and there is an edge $e \in E$ between vertex $v_i$ and its neighbour vertex $v_j$. It consists of two phases i.e., determining the originator nodes $O = \{O_1, O_2, ..., O_q\}$ then followed by simulating weighted message flows from originator nodes to their corresponding neighbour nodes. The weight of the message is the estimate of probability of reaching any node from originator node $O_l$ i.e., $M_{weight} = \frac{1}{Degree(O_l)}$. A recipient node maintains cumulative weights $TotalWeight$ of messages it received, corresponding to the message originator. The duration of simulation is controlled by a predefined Time-to-Live ($TTL$ for short) parameter. When a message arrived at vertex $v_i$, prior of re-circulation to the vertex's neighbour, the algorithm decrements the value of $TTL$ by 1 and divides the $M_{weight}$ by the degree of $v_i$. At the end of the simulation (i.e., $TTL$ is 0 or $M_{weight}$ becomes insignificantly low), each node sorts the $TotalWeight$ by originator then joins the cluster led by the originator having maximum $TotalWeight$. Figure \ref{fig:clustering} illustrates the use of the CDC clustering method applied to an overlay network. 

Intuitively, the quality of resulting clusters is highly dependent on the selection of ``good" originator nodes. The CDC method addresses this problem by defining criteria of a good originator, one of which is that a node is considered as good originator if the total weight it acquired mainly contributed by the message originated from itself rather than those originated from other originator. To satisfy this criteria, CDC adopts Two Hop Return Probability (THP) which is an approximation technique that determines the probability of returning to a node $v_l$ in the graph in two hops when a random walk is performed starting at $v_l$.

\section{Residual Performance-based Data Placement}\label{sec:design}

This section describes the design of the proposed Residual Performance-based Data Placement (RPDP) method. We first formulate the problem statement and the key assumptions in subsection \ref{subsec:design_problem}, which then followed by an overview of the design solution in subsection \ref{subsec:design_solution}. The remaining subsections provide a detailed description of the solution.

\subsection{Problem statement and assumptions}\label{subsec:design_problem}

A P2P storage system consists of $n$ heterogeneous data nodes distributed geographically with various network distances, denoted by $\mathcal{S} = \{s_1, s_2, ..., s_n\}$. The nodes in the system are connected to peers (neighbours) using Kademlia overlay network. Assume that we have data items $\mathcal{D} = \{d_1, d_2, ..., d_c\}$ to be stored in the P2P storage system with $r$ replication factor (i.e., using classic $n$-way replication). The goal of the data placement method is to select the most appropriate node $s_i$ to store primary replica of data $d_i$ in such a way that can balance the workload of the nodes in respect to their individual performance, leading to improved overall latency of the system. Further, to minimize the impact to the scalability of the P2P storage system, the data $d_i$ stored using this placement method should be retrievable by any nodes in the system without the need of global knowledge. The key assumptions used are: (i) the data size should be uniform, therefore, prior of storing data into the P2P storage system, if the data size is greater than predefined threshold, it must be split into $c$ fixed-size chunks, (ii) data placement for the remaining $r-1$ replicas performed by repeating the same process. 

\subsection{Solution overview}\label{subsec:design_solution}

We present Residual Performance-based Data Placement (RPDP), a data placement method for P2P storage system that selects of most appropriate nodes in the system to store data based on their temporal residual performance. The overall latency in a P2P system is affected by nodes with slowest performance, also known as stragglers \cite{AlAbbasi2020}. Therefore, by balancing the workload of the nodes in respect to their individual performance can lead to improved overall latency of the system. {\it Residual performance} is defined as the remaining performance capacity of a node to handle incoming workload, measured using throughput and latency, two important performance metrics commonly used in any arbitrary storage systems \cite{Chen1993, Chen1994, Hafner2004}. Nodes in the P2P storage system can be generalized as an ordinary storage system, therefore such metrics can be applied. We argue that by placing data to nodes with higher residual performance will result in an improved workload balance across nodes because at any point of time, a new data will be placed in a node with the least workload.

The RPDP data placement method requires knowledge of performance status of all the nodes in the system, which is used for data placement process. To achieve this, some nodes are designated as monitor nodes which responsible in maintaining the consolidated periodic performance status metrics of all other nodes and periodically performs ranking that provides dynamic score-board like global view of residual performance of all nodes in the system. The monitor node also keeps track of latest global minimum and maximum value of throughput and latency which are needed by other nodes to calculate their periodic local residual performance. To minimize the impact to the scalability and reduce communication overheads, the network is partitioned into multiple clusters where for each cluster, one of the nodes is selected to be the monitor node and the remaining ones designated as data nodes. The data nodes only need to provide their status metrics to the monitor node within their cluster. Subsection \ref{subsec:design_clustering} provides detail description of the network clustering while process of periodic status metrics update is detailed in subsection \ref{subsec:design_status_update}. 

To perform data placement, RPDP retrieves the most appropriate data nodes from monitor node (nodes with highest rank in the score board) then store the data at those nodes. The nodes in P2P storage system considered in this work are connected to peers using Kademlia overlay network and the data retrieval method corresponds to the RPDP data placement method relies on this network and its protocol i.e., using Kademlia-based DHT. RPDP method introduces modification to the DHT data structure that allows stored data retrieved without prior knowledge of the data location, eliminating the need for central components. Subsection \ref{subsec:design_data_placement} describes the detail of data placement process while subsection \ref{subsec:design_data_retrieval} provides detailed description of how data retrieval works with the modified DHT data structure.

\subsection{Network clustering}\label{subsec:design_clustering}

The RPDP data placement method requires designation of a certain node to perform nodes selection among all the other nodes to store data upon receiving a data placement request. Therefore, we first introduce different roles of a node for the P2P storage system i.e., {\it data node} and {\it monitor node}. The former is responsible for the date storage and periodically provide sits current status metrics to the monitor node. The monitor node in turn consolidates the status metrics of all data nodes and uses the information for the data nodes selection. Having a single monitor node across the P2P storage system can incur significant communication overhead with data nodes which increases network bandwidth consumption, especially if the nodes are further apart. As solution to this problem, the P2P storage system is partitioned into multiple clusters using the CDC clustering method (details of this method provided in Section \ref{sec:background}). We selected this clustering method because it is a distance-based method that groups together nodes that are in shortest distance from predefined originator nodes (i.e., cluster center), where the distance is measured by number of hops. After node clustering, for each cluster, one of the nodes is selected to be the monitor node while the remaining ones designated as data nodes. This approach can reduce the communication overhead between data nodes and the monitor node, because data nodes only need to provide their status metrics to the monitor node within their proximity i.e., within the same cluster.

\subsection{Status metrics update}\label{subsec:design_status_update}

RPDP requires the selection of the most appropriate data nodes to store data based on their temporal residual performance. The method considers three important performance metrics during the selection, namely throughput, latency, and storage space. Latency is defined as the time taken to respond to an I/O request (typically in seconds), while throughput is the amount of data (typically in megabytes or kilobytes) that can be read or written through the system per unit of time (typically in seconds). Higher workload generates a higher system utilization (e.g., CPU, RAM, disk) and increases throughput \cite{Chen1993}. If the system has not reached its maximum throughput, the average latency does not much affected with the increasing workload. At this point of time, it is considered that the system  has still sufficient residual performance. However, at the point when the system reaches its maximum throughput, the new incoming requests will be experiencing longer waiting time and therefore the average latency will increase. At this instance, it is considered that the system has lesser residual performance. To this end, it can be determined whether a data node is under higher workload if it is having high latency and high or reaching maximum throughput.

\begin{figure}[htbp]
\centerline{\includegraphics[width=0.5\textwidth]{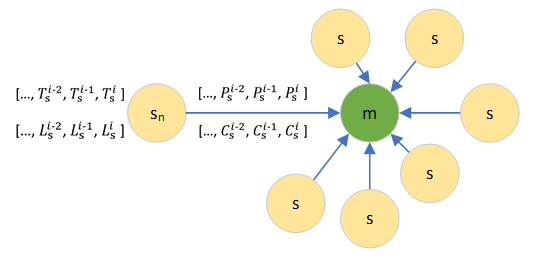}}
\caption{Periodic Status Metrics Update}
\label{fig:status_update}
\end{figure}

Each data node, say $s$, maintains historical data of its periodic average throughput, denoted by $T_s = \{..., t_{i-2}, t_{i-1}, t_i\}$ and periodic average latency, denoted by $L_s = \{..., l_{i-2}, l_{i-1}, l_i\}$, for sequential time period $i$. To minimize storage overhead, only limited number (system parameter) of most recent history entries are maintained. At the end of any given time period, each data node calculates its residual performance $P_s^i$ including the residual value of storage space $C_s^i$ (i.e., the remaining available storage space) and sends this information to the monitor node, as shown in Figure \ref{fig:status_update}.

For each request received within time period $i$, upon completing processing of a request, data nodes calculate the corresponding throughput $t(i,j)$ (in megabytes per second) and latency $l(i,j)$ (in seconds) and assigned a timestamp. The latency calculation includes the wait-time of the requests in the queue but excludes the transmission-time i.e., the time needed for a request to arrive at the queue of the data node. At the end of time period $i$, the average throughput $T_s^i$ and average latency $L_s^i$ are calculated by considering all completed requests received within the period, as shown in (\ref{eq:throughput}).

\begin{equation}\label{eq:throughput}
T_s^i = \frac{1}{|t(i,j)|} \sum_{j=1}^{|t(i,j)|} t(i,j), \quad L_s^i = \frac{1}{|l(i,j)|} \sum_{j=1}^{|l(i,j)|} l(i,j)
\end{equation}

 The next step is to compute the residual values for all the performance metrics considered. For the average throughput and average latency, because their values are in different unit and scale, they are first need to be normalized. Each data node first retrieves the latest global minimum and maximum value of throughput ($T_{min}$, $T_{max}$) and latency ($L_{min}$, $L_{max}$) from monitor node, to be used in the normalization of local average throughput and average latency. We select the min-max normalization method because it is suitable for the type of the data and it is one of the common data normalization methods. This method transforms the data into value in the range of 0 and 1, by transforming minimum value of data into a 0, maximum value into a 1, and every other values into a decimal between 0 and 1. Residual value for each performance metric is then calculated by subtracting the value from 1. As for calculation for the residual storage space (i.e., the remaining available storage space) for the period, $C_s^i$ is self explanatory. Residual value for storage space does not need to normalized because it only being used as a filtering criteria in the selection of data nodes.

\begin{equation}\label{eq:rp_throughput}
\Acute{T}_s^i = 1 - (\frac{T_s^i - T_{min}} {T_{max} - T_{min}})
\end{equation}

\begin{equation}\label{eq:rp_latency}
\Acute{L}_s^i = 1 - (\frac{L_s^i - L_{min}} {L_{max} - L_{min}})
\end{equation}

The residual values of average throughput and average latency are used as input to calculate the residual performance $P_s^i$ of a data node for duration $i$. The $P_s^i$ aggregates the inputs using average method. Coefficients $\omega_1$ and $\omega_2$ are assigned to both input metrics which can be used to fine-tune or adjust their weight or importance. 

\begin{equation}\label{eq:rp_aggregate}
P_s^i = \frac{1}{2}(\omega_1 \Acute{T}_s^i + \omega_2 \Acute{L}_s^i), \quad \omega_1 \leq 1, \omega_2 \leq 1
\end{equation}

\subsection{Data placement}\label{subsec:design_data_placement}

\begin{figure*}[!h]
\centerline{\includegraphics[width=0.75\textwidth]{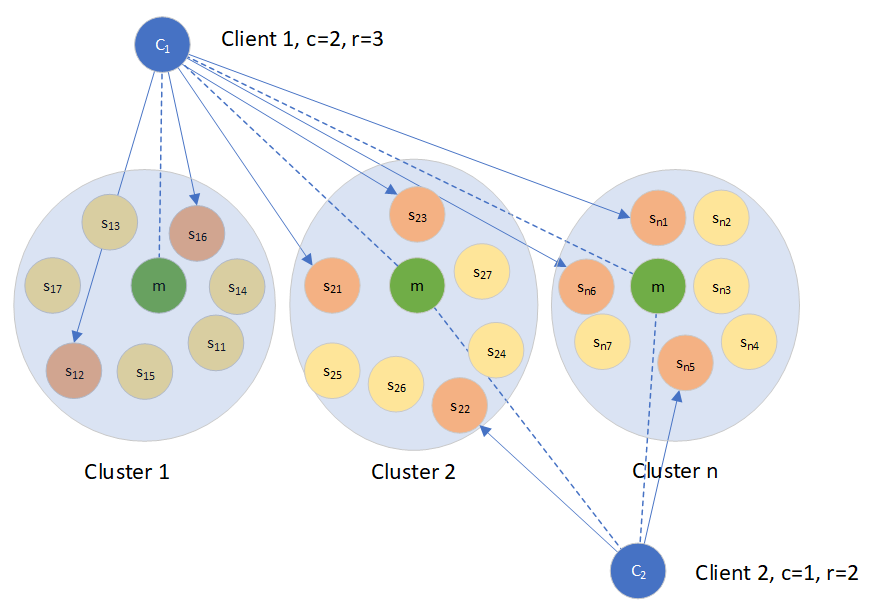}}
\caption{Replica Placement}
\label{fig:replica_placement}
\end{figure*}

As mentioned earlier, RPDP aims to balance the workload across all the nodes with respect to their individual performance to improve the overall latency. It selects the most appropriate data node to store data by considering the temporal status of all the nodes within the selected cluster. More specifically, given an input $c$, the method selects top-$c$ data nodes with sufficient remaining storage space and highest residual performance. The classic $n$-way replication is selected as the data availability approach. The number of replicas, $r$, durability, among other rules, is determined by the client application as part of the data storage policy. It is assumed that value of $r$ has an upper bound which is equal to the number clusters in the P2P storage system. As shown in Figure \ref{fig:replica_placement}, to store $c$ data with $r$ replication, given a list of monitor nodes, provided by the P2P storage system, the client selects the closest one from itself to request for $c$ data nodes to store $c$ data within the cluster. The method to determine the network distance between the client and the monitor node is considered beyond the scope of this work. However, one simplest method that can be used is for the client to perform network routing test for each of the candidate monitor nodes and select the one with the least number of network hops. Subsequently, for the remaining $r-1$ replicas, to ensure load balance across clusters, it is assumed the client will randomly choose $r-1$ monitor nodes and perform similar data store operation. 


Throughput calculation is affected by the size of the data, which introduces a constraint i.e., the data size should be uniform, hence the key assumption in Subsection \ref{subsec:design_problem}. To store data, client selects and connects to a monitor node to request for most suitable data node to store the data into. If the data is split into multiple chunks, the request will be for a set of data nodes. The client also includes the size of the data when sending the request to the monitor node. Recall from subsection \ref{subsec:design_status_update} that each data node needs to periodically send its status metrics (current residual performance and remaining available storage space) to the monitor node within the same cluster. Upon requested, using the status metrics of the data nodes and the attributes of the data to be stored, the monitor node first filters out the data nodes that do not have sufficient storage space. As shown in (\ref{eq:s_best}), on the remaining data nodes, it selects the top-$k$ data nodes with highest residual performance and return the result list, $S_{best}$ to the client.

\begin{equation}\label{eq:s_best}
S_{best} = top_c\{s_1, s_2, ..., s_n\} 
\end{equation}

The data node selection in RPDP is based on the temporal status of the data nodes, as such, it is improbable that the ID of the selected data node is closest (by XOR-distance of ID) to the data. Therefore, to ensure the data can be placed into and subsequently be retrieved from the selected data node, some modifications to data store protocol and DHT data structure of a typical Kademlia implementation is necessary. It is important to note that for RPDP, we assume that data will only be placed at a single data node whose ID is closest to the ID of the data, and that the client implements an audit function to ensure availability and durability of stored data. As such, RPDP does not implement data republication related functions. We have excluded here the details of such audit function.

\begin{figure*}[!h]
\centerline{\includegraphics[width=0.75\textwidth]{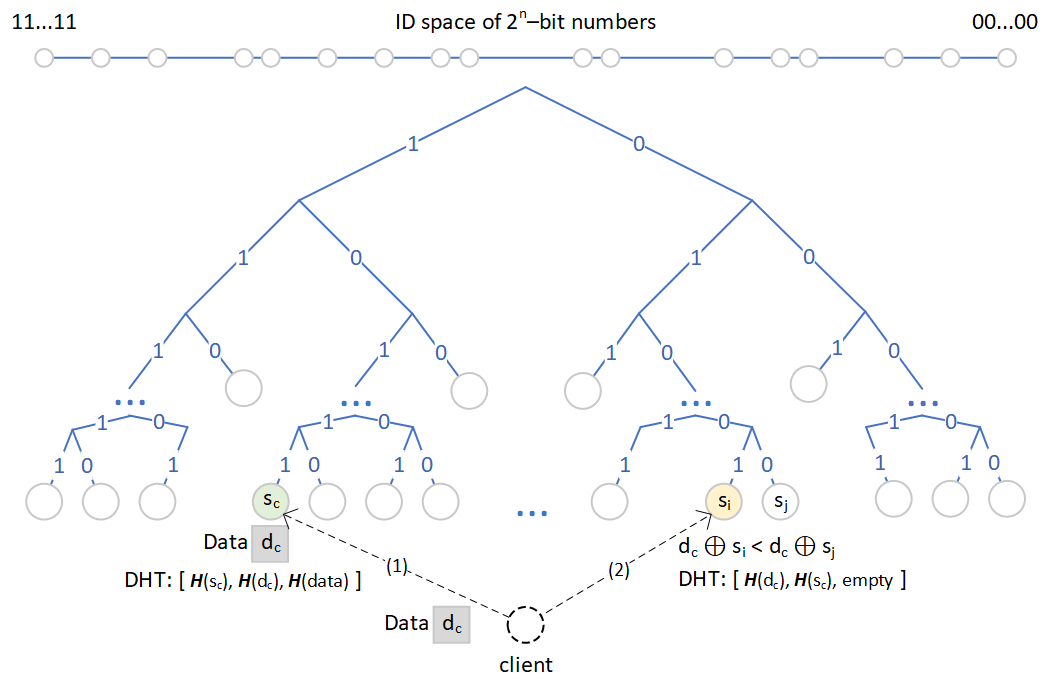}}
\caption{Data Placement. $d_c$=data, $s_c$=actual data node, $s_i$=mapping data node}
\label{fig:data_placement}
\end{figure*}

Prior to further explanation, we first establish a clear distinction between the definition of $key$ and ID in P2P systems which frequently used in the (Kademlia) literature. Every node in Kademlia P2P network has an opaque ID which typically assigned randomly when they join the network by running a hash function $H$ that generates a unique identifier within a predefined identifier space e.g., 160-bit SHA-1 or some larger space such as 256-bit SHA-2. For example, in P2P system using 256-bit identifier space, possible value of ID of a node is any integer between 0 and $2^{255}$. Data is stored in $\langle key, value \rangle$ pair format in Distributed Hash Table (DHT) structure, where the key is also taken from the same key space. How to derive the key for the data is not defined in the Kademlia protocol, however, it is typically derived by running the same hash function $H$ taking value of the data as the input \cite{Daniel2022}. The $\langle key, value \rangle$ data is stored at a node with ID closest to the key of the data which determined by XOR-based distance between the node ID and the key of the data \cite{Maymounkov2002}. The use of terminology ID for node and $key$ for data can sometimes create perplexities, therefore in this paper, we use ID to denote identifier from key space for both node and data. As such, format of stored data is now represented as $\langle id, value \rangle$.

The RPDP data placement method is a two-fold process. In the first step, after retrieving the information of selected data nodes from monitor node, the method directly perform data store to the node, followed by second step in which the algorithm proceeds looking up for XOR-based closest node to the data to store a mapping between the data and its actual location. The detail description of each steps are as follows:

\paragraph{Step 1} As shown in Figure \ref{fig:data_placement}, the client directly connects and stores data into the selected node $s_c$. This creates a problem where $s_c$ is not necessarily and even unlikely to be the closest node to $d_c$. As a solution to this problem, RPDP introduces the concepts of virtual node, actual node, primary ID, and secondary ID. Assuming that $s_i$ is the closest node to $d_c$, then the data is supposedly stored at $s_i$. With RPDP, the data is stored at the selected node $s_c$, instead. From the perspective of $d_c$, RPDP refers to $s_i$ as the virtual node, $s_c$ as the actual node, and all the other nodes as normal node. The primary ID is the ID that being used during data look up at normal and virtual nodes while the secondary ID is being used in conjunction with primary ID to further identify the data stored in the actual node. In order to store $d_c$ to the actual node $s_c$ without breaking the fundamental rule of closeness in Kademlia, its primary ID $d_c^p$ must be closer to $s_c$ compared to the distance between $s_c$ to its nearest neighbour, say $s_l$. Formally, this constraint is defined as $d_c^p \oplus H(s_c) < H(s_c) \oplus H(s_l)$. To satisfy such constraint, RPDP selects the simplest solution which is by assigning primary ID of $d_c$ equal to the ID of the actual node $s_c$ and its secondary ID $d_c^s$ equal to the ID of the data itself. Formally, $d_c^p = H(s_c), d_c^s = H(d_c)$.

\begin{figure}
\centerline{\includegraphics[width=0.45\textwidth]{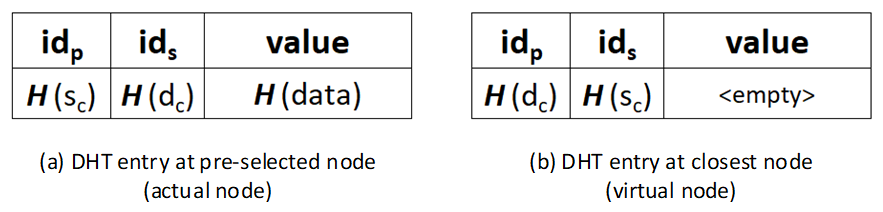}}
\caption{DHT Entry}
\label{fig:dht_entry}
\end{figure}

Having primary and secondary ID to the data means each data node now maintains list of triples $\langle id_p, id_s, value \rangle$ for all data it stores. Consequently, a single DHT entry of data in the node is now can only be uniquely identified and retrieved by using both of primary and secondary ID. After the step 1 in RPDP, the actual node (the selected node) $s_c$ will have a DHT data entry of $\langle H(s_c), H(d_c), value \rangle$, as shown in Figure \ref{fig:dht_entry}(a). 

\begin{figure*}[!h]
\centerline{\includegraphics[width=0.75\textwidth]{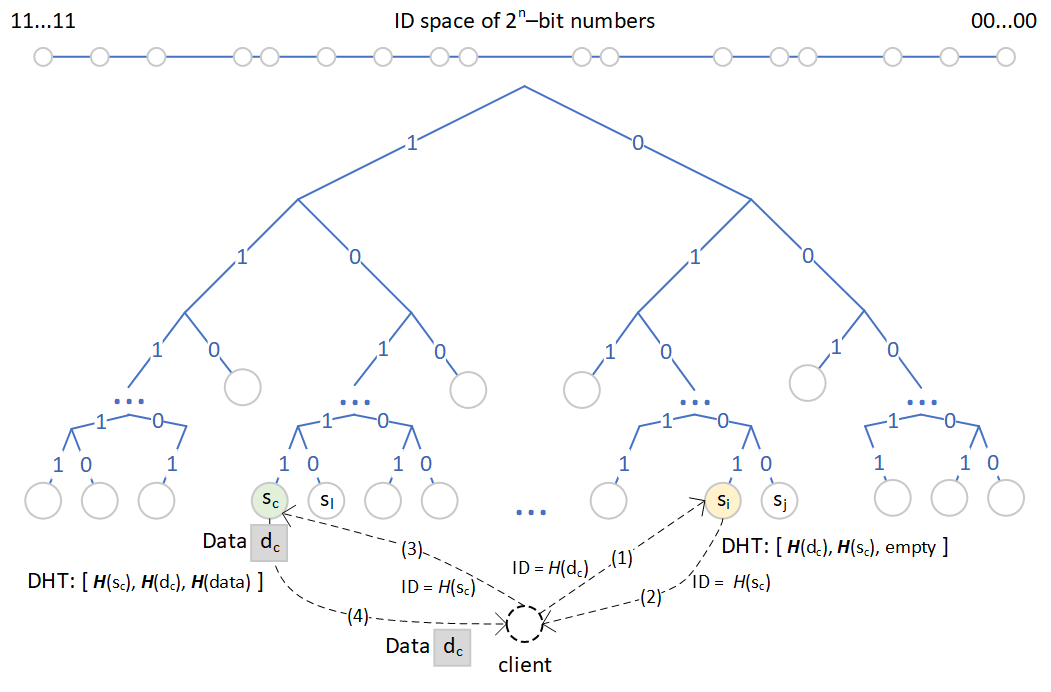}}
\caption{Data Retrieval. $d_c$=data, $s_c$=actual data node, $s_i$=mapping data node}
\label{fig:data_retrieval}
\end{figure*}

\paragraph{Step 2} RPDP proceeds to look up for the closest node to $d_c$. Upon arrives at the closest node $s_i$ (i.e., the virtual node), the algorithm assigns primary ID of $d_c$ equal to the ID of the data itself and its secondary ID $d_c^s$ equal to the ID of node where the data is actually stored $s_c$ (i.e., the actual node). Formally, $d_c^p = H(d_c), d_c^s = H(s_c)$. At the virtual node $s_i$, the algorithm creates a triples $\langle H(d_c), H(s_c), value=empty \rangle$ DHT entry to serve as mapping between primary ID of the data to its secondary ID, as shown in Figure \ref{fig:dht_entry}(b). It is important to emphasize that $s_i$ does not store the actual data, hence, $value = empty$. Instead, it provides a mechanism to ensure that with given ID of the data, using a typical Kademlia look up, RPDP algorithm will be able to return the information of the actual node ($s_i$). Further explanation on this mechanism is elaborated in the next sub section.


\subsection{Data retrieval}\label{subsec:design_data_retrieval}

To look up data in a typical Kademlia implementation, a node checks through its DHT by using the ID of the data as the key. If an entry exists, the value corresponding to the ID is returned, otherwise, the node performs node look up and returns the closest node to the ID of the data that it knows of (from its $k$-buckets). RPDP data placement method introduces a modification to the DHT data structure in which value at actual node is identified by using both primary ID and secondary ID, therefore we have devised a simple data retrieval method accordingly. In this method, a node first checks the $id_p$ in the DHT using the ID of data, If the entry exists, the algorithm branches out to two possibilities: (1) there are no values associated to the ID i.e., $value = empty$; or (2) there are one or more value entries associated to the ID. In the scenario of first possibility, the node treats the entry as a map that points to the secondary ID $id_s$. The node then proceeds to perform further data look up but this time is using the value of $id_s$. The look up eventually arrives at the node closest to the $id_s$ and at this point, the method goes into a scenario corresponds to the second possibility above. In this scenario, the node further filters the DHT entries by selecting the one having $id_s$ equal to the ID of the data itself. Figure \ref{fig:data_retrieval} shows the illustration of this process.


For simplicity, we omitted the data caching and optimisation techniques for the data retrieval method.

\section{Evaluation}\label{sec:evaluation}

We implemented the RPDP data placement method and conducted experiments using PeerSim \cite{Montresor2009}, a highly scalable P2P simulator written in Java programming language capable of simulating P2P systems with a network consisting of millions of nodes. As the baseline system to compare with, we also implemented the state-of-the-art Kademlia-based DHT data placement method. Experiments were conducted to measure the overall latency and latency variance of the systems under continued workload. We also experimented the scenario of growing number of nodes to observe the scalability of the proposed RPDP method. Lastly we provided an analysis of the additional complexities of the corresponding data retrieval method as the result of the RPDP implementation.

\subsection{Experimental setup}\label{subsec:eval_setup}
The simulation environment using PeerSim ran on a single machine with Microsoft Windows 10 Enterprise operating system, having 11th Gen Intel Core i5 2.6GHz 4 cores CPU, 8 Gigabytes of memory, and 500 Gigabytes of storage device. The network was established by creating predefined number of nodes (input parameter) and assigning each node with random identifier taken from 160-bit ID space and allocating random bootstrap nodes to each node that were used to establish the routing table during warm-up period. At this stage, each node also assigned a random number taken from standard normal distribution to represent heterogeneous maximum throughput across the nodes. By assuming every node operates at their maximum throughput, with fixed block size of data (set as 1 Megabytes), latency for each node can be calculated by dividing the block size with the maximum throughput. For simplicity, we also assumed that such latency is the same for both read and write operations. The warm-up process (i.e., process of random node look ups) was set to run for 1 hour to allow routing table ($k$-buckets) for each node sufficiently populated with peers/neighbours information. Specific for RPDP, the network then partitioned based on desired number of clusters (input parameter) and allocation of corresponding monitor nodes. At this stage, both systems (baseline and RPDP) are considered in initial state. Immediately following the initial state, was the running state, simulated by generating a number of data placement requests (input parameter) at randomly selected nodes at fixed predefined interval (i.e., every 1 second). Data for each operation was generated and assigned random ID taken from the same 160-bit ID space. The running state was set to run 1 second after the warm-up period until the end of simulation. To simplify the experiments, it was assumed that each node has sufficient storage device to store the simulated data. Few hours into the running state (set as 3 hours), the experiment periodically (e.g., every 1 minute) captured the latency of each node and calculated the overall (average) and variance (standard deviation) of latency of the simulated P2P systems. The overall time for simulation was set for 10 hours and assuming that all nodes remains active in the network throughout the simulation time i.e., scenario related to network churn were excluded from the experiment. 

\subsection{Overall system latency}\label{subsec:eval_latency}
\begin{figure}
  \begin{center}
    \subfigure[Overall Latency (Kademlia-based DHT)]{
      \includegraphics[width=0.48\linewidth]{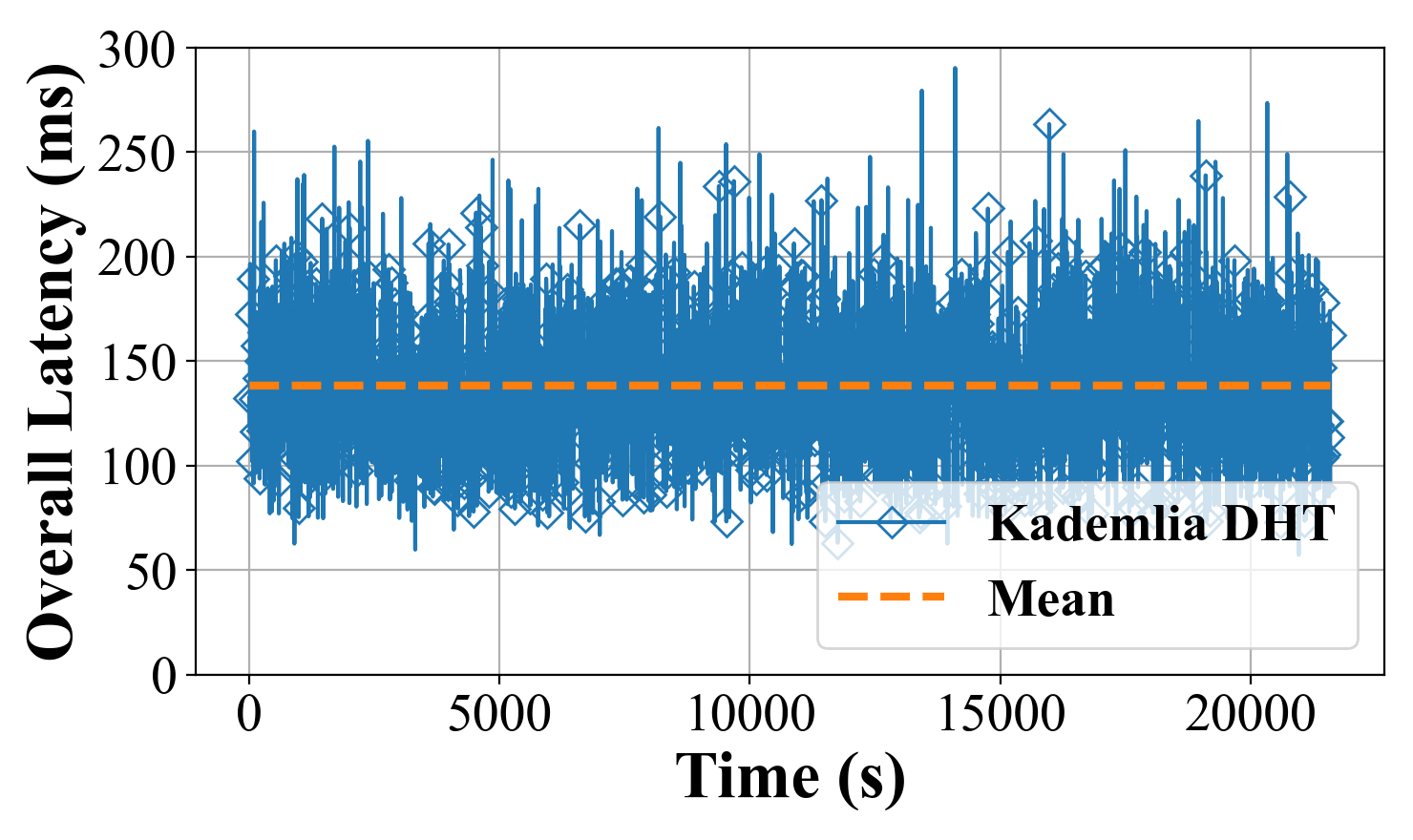}}
    \subfigure[Overall Latency (RPDP)]{
      \includegraphics[width=0.48\linewidth]{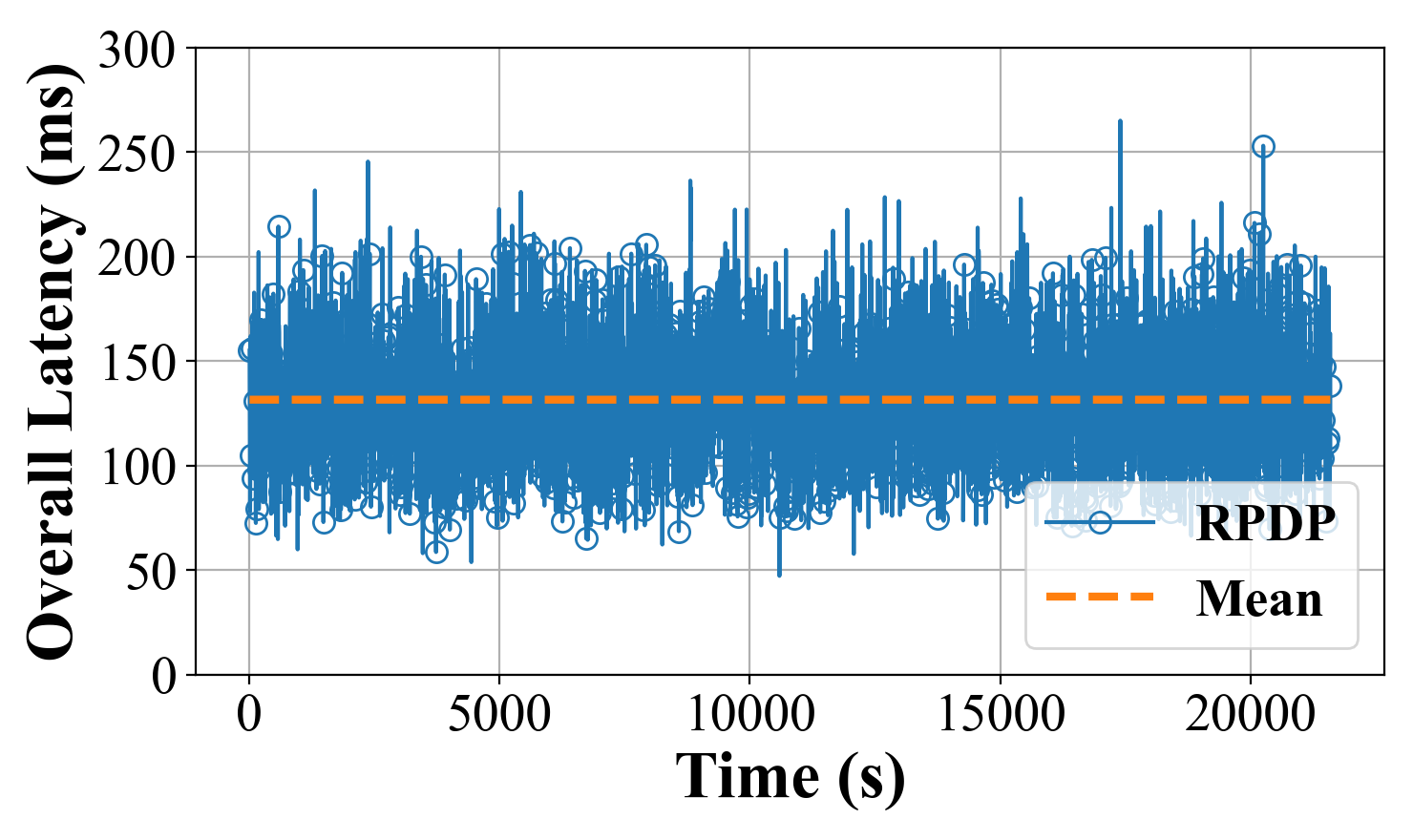}}
  \end{center}
  \caption{Overall latency}
  \label{fig:evaluation_overall_latency}
\end{figure}

We analyzed and compared the overall latency of the P2P system that included the RPDP data placement method and compared it with baseline system that uses the state-of-the-art Kademlia-based DHT data placement. Figure \ref{fig:evaluation_overall_latency} shows the overall latency for both systems, based on simulation with 100 nodes, continuous fixed workload, and with statistic data capture of every 5 seconds. Although the difference is subtle, it can be seen that RPDP system (Figure \ref{fig:evaluation_overall_latency})(b) achieved lower overall latency compared to the Kademlia-based DHT (baseline) system (Figure \ref{fig:evaluation_overall_latency})(a). In particular for this scenario, RPDP achieves overall latency of 131.6 milliseconds while Kademlia-based DHT overall latency is 138.33 milliseconds, improved about 4.87\%.

\begin{figure}
  \begin{center}
    \subfigure[Latency Variance (Kademlia-based DHT)]{
      \includegraphics[width=0.48\linewidth]{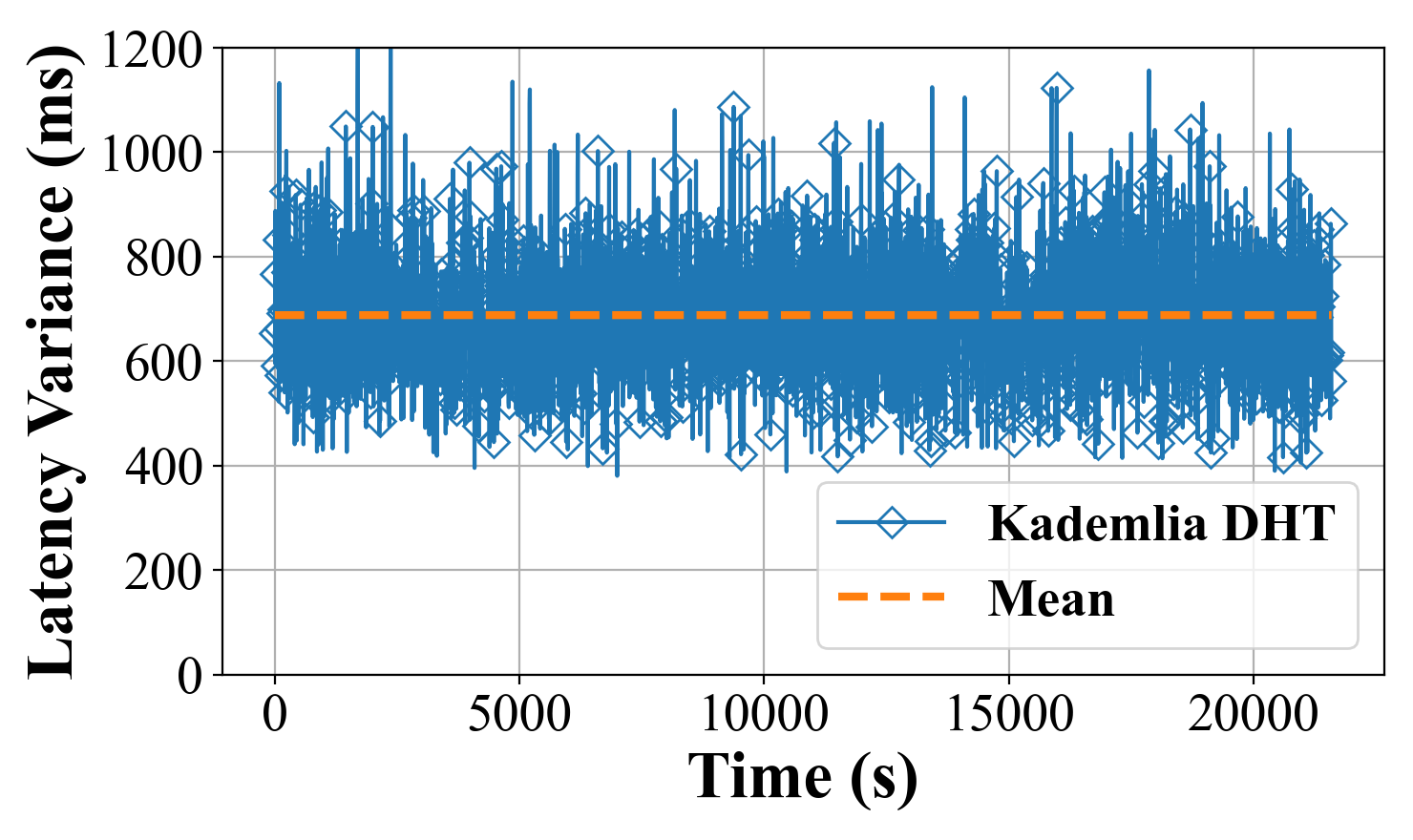}}
    \subfigure[Latency Variance (RPDP)]{
      \includegraphics[width=0.48\linewidth]{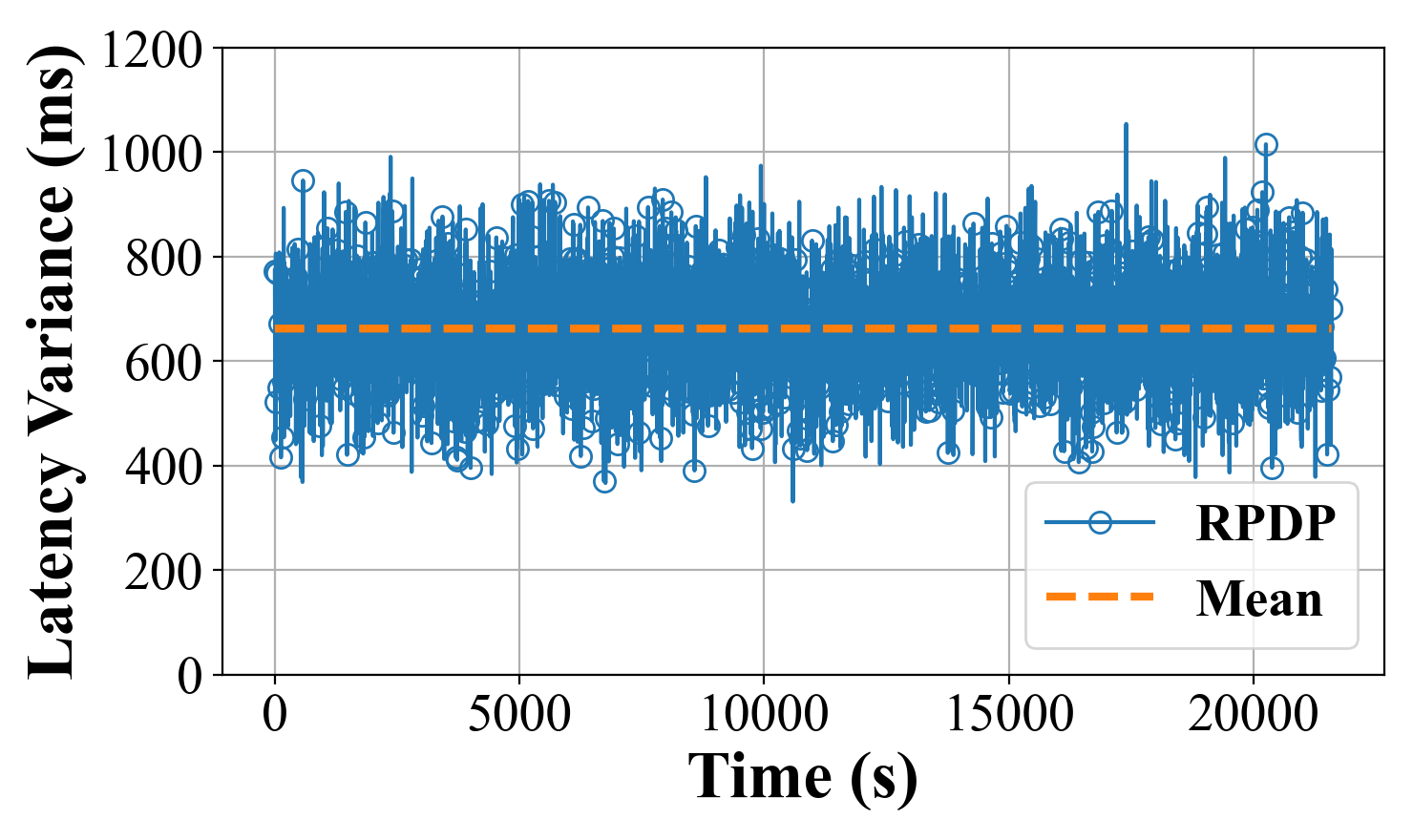}}
  \end{center}
  \caption{Latency variance}
  \label{fig:evaluation_latency_variance}
\end{figure}

The variance of overall latency for RPDP, as shown in Figure \ref{fig:evaluation_latency_variance}(a), is relatively less brittle compared to the Kademlia-based DHT shown in \ref{fig:evaluation_latency_variance}(b), indicating that distribution of nodes workload in RPDP is more balanced compared to the Kademlia-based DHT. The latter method places data at node based on a consistent hashing function irrespective to the performance of individual nodes leading to imbalanced load distribution. RPDP on the other hand achieves a more balanced load distribution by placing data based on residual performance of each node which in turn, results in lower overall system latency.


\subsection{Scalability}\label{subsec:eval_scalability}
\begin{figure*}[!h]
  \begin{center}
    \subfigure[Overall Latency]{
      \includegraphics[width=0.37\linewidth]{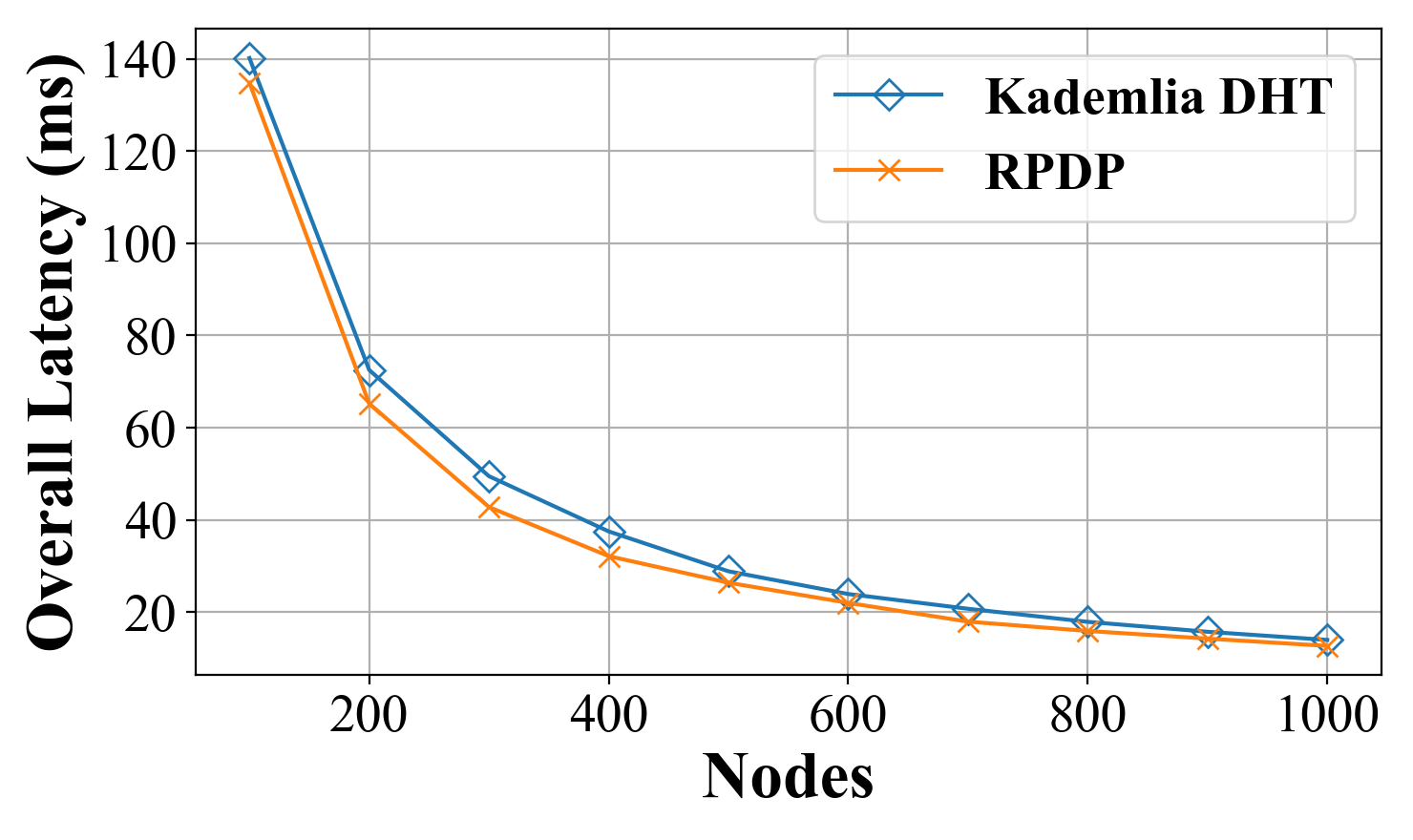}}
    \subfigure[Latency Variance]{
      \includegraphics[width=0.37\linewidth]{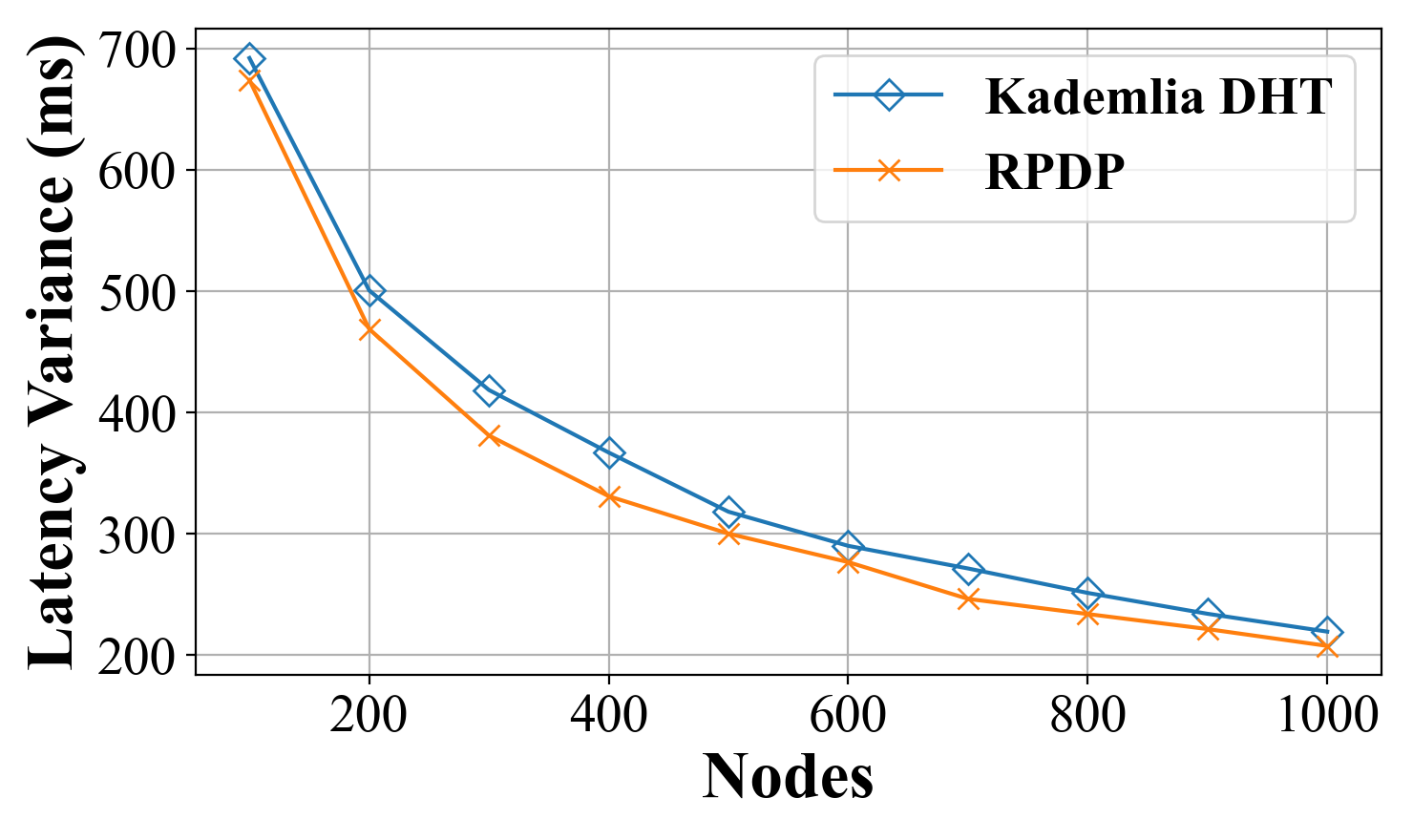}}
  \end{center}
  \caption{Overall latency and latency variance with growing number of nodes}
  \label{fig:evaluation_scalability}
\end{figure*}

We also conducted experiments to observe the overall latency  for both Kademlia-based DHT (baseline) and the RPDP system under growing number of nodes. Figure \ref{fig:evaluation_scalability} shows the results of the experiment using constant workload and varying number of nodes. It can be observed that the overall latency for both systems follow the similar decreasing trend due to the constant workload which demonstrates that the systems are scalable i.e., inversely, with increased workload, overall latency can be maintained by having proportionally more nodes in the network. More importantly, the results reveal that compared to the Kademlia-based DHT, RPDP method consistently achieve lower overall latency and latency variance for any given number of nodes. The curves appear to converge as the number of nodes grows in the network because the overall latency difference is getting smaller and approaching zero due to the constant workload.


\subsection{Data retrieval complexity}\label{subsec:eval_complexity}

The RPDP data placement method achieves lower overall latency and scalability at the expense of additional complexity during data retrieval. Recall from Subsection \ref{subsec:design_data_placement}, the method is implemented as two steps protocol i.e., data store at pre-selected data nodes (actual nodes) followed by creating and storing data map at the virtual nodes. Consequently, data retrieval needs to perform at most two node lookup operations i.e., first lookup to find the data map at virtual nodes, with $\mathcal{O}(\log{n})$ complexity, then followed by the second one to find the actual nodes based on data map. The complexity of the second lookup operation is the same as the first one therefore, it can be concluded that the complexity of data retrieval remains as $\mathcal{O}(\log{n})$. 

\section{Conclusion}\label{sec:conclusion}

Kademlia-based DHT data placement method has a limitation whereby it does not consider the dynamic performance of nodes which often leads to imbalance resource usage and therefore affecting the overall latency of the system. 
This paper described the RPDP method which offers an approach of selecting most appropriate data nodes based on their residual performance and relies on Kademlia-based DHT with modified data structure to allow data subsequently retrieved without the need of global knowledge. Compared to the Kademlia-based DHT, the RPDP method improves the overall latency of the system with minimal impact to the data retrieval complexity. Future line of work will include: (i) exploring the fairness of data distribution, (ii) the role of proxy node to perform the load balancing inter-cluster, (iii) non-primary replica placement based on data popularity, and (iv) migration of infrequently accessed data to the cold network partition implementing different coding mechanism e.g., erasure coding.

\bibliographystyle{IEEEtran}
\bibliography{conference_101719.bib}

\begin{thebibliography}{10}
\providecommand{\url}[1]{#1}
\csname url@samestyle\endcsname
\providecommand{\newblock}{\relax}
\providecommand{\bibinfo}[2]{#2}
\providecommand{\BIBentrySTDinterwordspacing}{\spaceskip=0pt\relax}
\providecommand{\BIBentryALTinterwordstretchfactor}{4}
\providecommand{\BIBentryALTinterwordspacing}{\spaceskip=\fontdimen2\font plus
\BIBentryALTinterwordstretchfactor\fontdimen3\font minus
  \fontdimen4\font\relax}
\providecommand{\BIBforeignlanguage}[2]{{%
\expandafter\ifx\csname l@#1\endcsname\relax
\typeout{** WARNING: IEEEtran.bst: No hyphenation pattern has been}%
\typeout{** loaded for the language `#1'. Using the pattern for}%
\typeout{** the default language instead.}%
\else
\language=\csname l@#1\endcsname
\fi
#2}}
\providecommand{\BIBdecl}{\relax}
\BIBdecl

\bibitem{Kermarrec2015}
A.-M. Kermarrec and F.~Ta{\"{i}}ani, ``{Want to scale in centralized systems?
  Think P2P},'' \emph{Journal of Internet Services and Applications}, vol.~6,
  no.~1, p.~16, Aug 2015.

\bibitem{Maymounkov2002}
P.~Maymounkov and D.~Mazieres, ``{Kademlia: A peer-to-peer information system
  based on the XOR metric},'' in \emph{International Workshop on Peer-to-Peer
  Systems}.\hskip 1em plus 0.5em minus 0.4em\relax Springer, 2002, pp. 53--65.

\bibitem{Lu2022}
K.~Lu, N.~Zhao, J.~Wan, C.~Fei, W.~Zhao, and T.~Deng, ``{RLRP: High-efficient
  data placement with reinforcement learning for modern distributed storage
  systems},'' in \emph{Proc. of IEEE International Parallel and Distributed
  Processing Symposium (IPDPS)}, May 2022, pp. 595--605.

\bibitem{Daniel2022}
E.~Daniel and F.~Tschorsch, ``{IPFS and Friends: A qualitative comparison of
  next generation peer-to-peer data networks},'' \emph{IEEE Communications
  Surveys \& Tutorials}, vol.~24, no.~1, pp. 31--52, 2022.

\bibitem{Trautwein2022}
D.~Trautwein, A.~Raman, G.~Tyson, I.~Castro, W.~Scott, M.~Schubotz, B.~Gipp,
  and Y.~Psaras, ``{Design and evaluation of IPFS: A storage layer for the
  decentralized web},'' in \emph{Proceedings of the ACM SIGCOMM 2022
  Conference}, Aug 2022, pp. 739--752.

\bibitem{Swarm2021}
\BIBentryALTinterwordspacing
``{SWARM: Storage and communication infrastructure for a self-sovereign digital
  society},'' 2012. [Online]. Available:
  \url{https://www.ethswarm.org/swarm-whitepaper.pdf}
\BIBentrySTDinterwordspacing

\bibitem{Storj2018}
\BIBentryALTinterwordspacing
``{Storj: A decentralized cloud storage network framework},'' 2018. [Online].
  Available: \url{https://www.storj.io/storj.pdf}
\BIBentrySTDinterwordspacing

\bibitem{Zhou2019}
J.~Zhou, Y.~Chen, W.~Xie, D.~Dai, S.~He, and W.~Wang, ``{PRS: A
  pattern-directed replication scheme for heterogeneous object-based
  storage},'' \emph{IEEE Transactions on Computers}, vol.~69, no.~4, pp.
  591--605, 2019.

\bibitem{Benet2014}
\BIBentryALTinterwordspacing
J.~Benet, ``{IPFS - content addressed, versioned, P2P file fystem},''
  \emph{CoRR}, vol. abs/1407.3561, 2014. [Online]. Available:
  \url{http://arxiv.org/abs/1407.3561}
\BIBentrySTDinterwordspacing

\bibitem{Swarm2020}
\BIBentryALTinterwordspacing
V.~Trón, ``{The book of Swarm, V1.0},'' 2020. [Online]. Available:
  \url{https://www.ethswarm.org/The-Book-of-Swarm.pdf}
\BIBentrySTDinterwordspacing

\bibitem{Safe2014}
\BIBentryALTinterwordspacing
N.~Lambert and B.~Bollen, ``{The SAFE network a new decentralised internet},''
  2014. [Online]. Available:
  \url{https://docs.maidsafe.net/Whitepapers/pdf/TheSafeNetwork.pdf}
\BIBentrySTDinterwordspacing

\bibitem{Safe2018}
\BIBentryALTinterwordspacing
``{The SAFE network primer},'' 2021. [Online]. Available:
  \url{https://primer.safenetwork.org/}
\BIBentrySTDinterwordspacing

\bibitem{AuroraFS2022}
\BIBentryALTinterwordspacing
``{AuroraFS Whitepaper},'' 2022. [Online]. Available:
  \url{https://www.aufs.io/yveepoom/2022/01/Whitepaper.pdf}
\BIBentrySTDinterwordspacing

\bibitem{Datta2009}
A.~Datta, \emph{{Peer-to-peer storage}}.\hskip 1em plus 0.5em minus 0.4em\relax
  Springer, 2009, pp. 2075--2081.

\bibitem{Sammy2021}
S.~de~Figueiredo, A.~Madhusudan, V.~Reniers, S.~Nikova, and B.~Preneel,
  ``{Exploring the {Storj} network: A security analysis},'' in \emph{Proc. of
  the 36th Annual ACM Symposium on Applied Computing}, 2021, pp. 257--264.

\bibitem{Reed1960}
I.~S. Reed and G.~Solomon, ``Polynomial codes over certain finite fields,''
  \emph{SIAM Journal of the Society for Industrial and Applied Mathematics},
  vol.~8, no.~2, pp. 300--304, 1960.

\bibitem{Mnih2015}
V.~Mnih, K.~Kavukcuoglu, D.~Silver, A.~A. Rusu, J.~Veness, M.~G. Bellemare,
  A.~Graves, M.~Riedmiller, A.~K. Fidjeland, G.~Ostrovski \emph{et~al.},
  ``Human-level control through deep reinforcement learning,'' \emph{Nature},
  vol. 518, no. 7540, pp. 529--533, 2015.

\bibitem{Singh2007}
A.~Singh and M.~Haahr, ``{Decentralized clustering in pure P2P overlay networks
  using Schelling's model},'' in \emph{Proc. of the IEEE International
  Conference on Communications}, June 2007, pp. 1860--1866.

\bibitem{Ramaswamy2005}
{Lakshmish Ramaswamy}, B.~Gedik, and L.~Liu, ``{A distributed approach to node
  clustering in decentralized peer-to-peer networks},'' \emph{IEEE Transactions
  on Parallel and Distributed Systems}, vol.~16, no.~9, pp. 814--829, Sep 2005.

\bibitem{AlAbbasi2020}
A.~O. Al-Abbasi and V.~Aggarwal, ``{TTLCache: Taming latency in erasure-coded
  storage through TTL caching},'' \emph{IEEE Transactions on Network and
  Service Management}, vol.~17, no.~3, pp. 1582--1596, Sep 2020.

\bibitem{Chen1993}
P.~M. Chen and D.~A. Patterson, ``Storage performance-metrics and benchmarks,''
  \emph{Proceedings of the IEEE}, vol.~81, no.~8, pp. 1151--1165, 1993.

\bibitem{Chen1994}
P.~M. Chen, E.~K. Lee, G.~A. Gibson, R.~H. Katz, and D.~A. Patterson, ``{RAID:
  high-performance, reliable secondary storage},'' \emph{ACM Computing
  Surveys}, vol.~26, no.~2, pp. 145--185, Jun 1994.

\bibitem{Hafner2004}
J.~L. Hafner, V.~Deenadhayalan, T.~Kanungo, and K.~Rao, ``Performance metrics
  for erasure codes in storage systems,'' \emph{IBM Res. Rep. RJ}, vol. 10321,
  2004.

\bibitem{Montresor2009}
A.~Montresor and M.~Jelasity, ``Peersim: A scalable {P2P} simulator,'' in
  \emph{Proc.of the IEEE Ninth International Conference on Peer-to-Peer
  Computing}, 2009, pp. 99--100.

\end{thebibliography}

\end{document}